\documentclass[notitlepage,superscriptaddress,showpacs,nobalancelastpage,twocolumn,aps,prl]{revtex4}

\usepackage[english]{babel}
\RequirePackage[T1]{fontenc}
\RequirePackage{times} 

\usepackage{siunitx} 
\usepackage[italicdiff]{physics}
\usepackage{amsfonts}
\usepackage{subfigure}
\usepackage{amsmath}
\usepackage{amssymb}
\usepackage{graphicx,epstopdf}
\usepackage{array}
\usepackage{hyperref}
\usepackage{xcolor} 
\usepackage{my_symbols}
\usepackage{CJK}
\usepackage{bm}
\usepackage{verbatim}
\usepackage[normalem]{ulem}
\usepackage{pdfpages}

{\par}

\newcounter{mycomment}

\newcommand\rmv{\bgroup\markoverwith {\textcolor{red}{\rule[0.5ex]{2pt}{0.4pt}}}\ULon}

\begin{document}
\begin{CJK*}{UTF8}{gbsn} 
\title{Purely magnetic logic based on polarized spin waves}
\author{Weichao Yu (余伟超)}
\email{Present address: Institute for Materials Research, Tohoku University, Sendai, Japan}
\affiliation{Department of Physics and State Key Laboratory of Surface Physics, Fudan University, Shanghai 200433, China}
\author{Jin Lan (兰金)}
\email{Present address: Center of Joint Quantum Studies and Department of Physics, School of Science, Tianjin University, Tianjin 300350, China }
\affiliation{Department of Physics and State Key Laboratory of Surface Physics, Fudan University, Shanghai 200433, China}
\author{Jiang Xiao (萧江)}
\email[Corresponding author:~]{xiaojiang@fudan.edu.cn}
\affiliation{Department of Physics and State Key Laboratory of Surface Physics, Fudan University, Shanghai 200433, China}
\affiliation{Institute for Nanoelectronics Devices and Quantum Computing, Fudan University, Shanghai 200433, China}

\begin{abstract}

Spin wave, the precession of magnetic order in magnetic materials, is a collective excitation that carries spin angular momentum.
Similar to the acoustic or optical waves, the spin wave also possesses the polarization degree of freedom. Although such polarization degrees of freedom are frozen in ferromagnets, they are fully unlocked in antiferromagnets or ferrimagnets.
Here we introduce the concept of magnetic gating and demonstrate a spin wave analog of the Datta-Das spin transistor in antiferromagnet.
Utilizing the interplay between polarized spin wave and the antiferromagnetic domain walls, we propose a universal logic gate of pure magnetic nature, which realizes all Boolean operations in one single magnetic structure. We further construct a full functional 4-bit Arithmetic Logic Unit using only sixteen spin wave universal logic gates, operating in a weaving fashion as a Jacquard loom machine. The spin wave-based architecture proposed here also sets a model for the future energy efficient non-volatile computing, the distributed processing-in-memory computing, and the evolvable neuromorphic computing.
\end{abstract}

\maketitle
\end{CJK*}

{\it Introduction.}
At present, most of the developments in spintronics, including the discovery of giant magneto-resistance (GMR) \cite{baibich_giant_1988,binasch_enhanced_1989}, the spin-transfer torque (STT)
\cite{slonczewski_current-driven_1996,berger_emission_1996}, the spin-orbit torque (SOT) \cite{manchon_current-induced_2018}, and the invention of the STT(SOT)-MRAM and the magnetic racetrack memories \cite{parkin_magnetic_2008,parkin_memory_2015}, concentrate on data storage. And they all rely on the spin current carried by the conduction electrons, which give rise to the unavoidable Joule heating. To go beyond above limitations of spintronics, it is highly desirable to employ the types of spin carriers which not only dissipate less but can serve for the both purposes of data storage and processing. One of the promising candidates is the spin wave (or magnon) \cite{kajiwara_transmission_2010,chumak_magnon_2015},  the collective precession of ordered magnetization in magnetic materials.

For the purpose of data processing based on spin waves, most efforts so far have been using the spin wave amplitude or phase to encode information \cite{vogt_realization_2014,chumak_magnon_2014,vogt_realization_2014,lan_spin-wave_2015,klingler_spin-wave_2015}. However, it is more natural to use the more robust polarization degree of freedom to encode information \cite{lan_antiferromagnetic_2017}. In ferromagnet, the polarization freedom is frozen because only the right-circular polarization is allowed. In antiferromagnet, in contrast, both left- and right-circular polarizations exist and they are degenerated \cite{keffer_theory_1952,cheng_antiferromagnetic_2016,lan_antiferromagnetic_2017}. With two circular modes, spin waves possessing arbitrary polarizations can be constructed, just like its optical counterparts.

In this paper, we introduce the concept of magnetic gating, where the exchange field from a magnetic gating layer can shift the spin wave dispersions in an antiferromagnet. In the mean time, the degeneracy of the two circular spin wave modes is lifted by the gating magnetization, causing a polarization rotation for the linearly polarized spin waves. This magnetic gating effect on spin wave polarizations, together with the polarization-selective spin wave driven domain wall motion presented by the same authors previously \cite{yu_polarization-selective_2018}, forms a complete inter-conversion scheme between the static magnetic texture and dynamical spin waves via the polarization channel.
Based on this scheme, we propose a purely magnetic logic gate, whose inputs and outputs are both the non-volatile magnetic racetrack memories \cite{parkin_magnetic_2008,parkin_memory_2015}, and the information processing in between is accomplished by polarized spin waves. Superior to most existing architectures, such a magnetic logic gate is capable of achieving all unary and binary logic operations in one single hardware structure.

\begin{figure*}[t]
\centering
\includegraphics[width=0.99\textwidth, trim=0 35 0 15,clip]{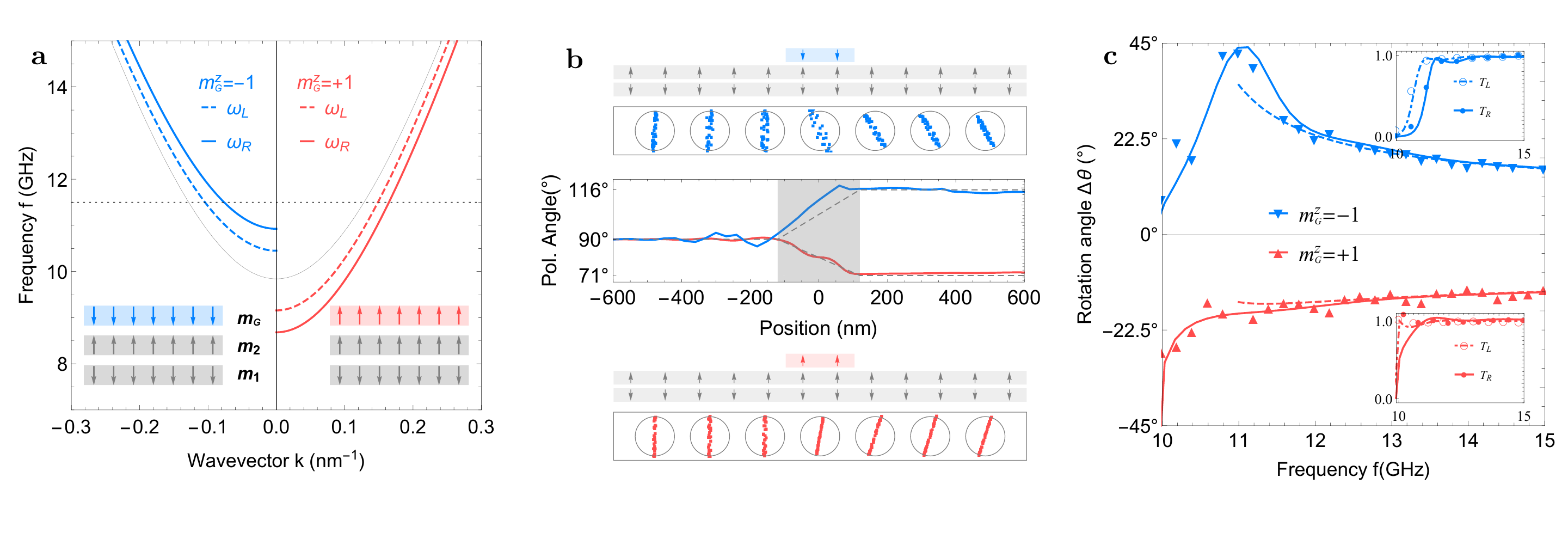}
\caption{Magnetic gating effect on spin wave polarization. (a) The spin wave dispersions for the left- and right-circular modes for ungated (gray), up-gated (red), and down-gated (blue) SyAF. Inset: an SyAF gated magnetically by the capping layer $\mb_\ssf{G}$.
(b) The spatial variation of the spin wave polarization across an ungated-gated-ungated SyAF junction extracted from the micromagnetic simulations. The vertically polarized spin wave of frequency $f=\SI{11.5}{GHz}$ is injected from the left side.
Top/Bottom: The Lissajous-like pattern showing the stagger order $n^{x,y}\equiv(m^{x,y}_2-m^{x,y}_1)/2$ for the down-/up-gated case.
Middle: The polarization angle as function of position.
(c) The spin wave polarization rotation angle $\Delta\theta$ as function of spin wave frequency for a down-/up-gated SyAF. The data points are extracted from micromagnetic simulations, and the curves are calculated from a full scattering calculation (solid) and the WKB approximation (dashed). Insets: the transmission probability across the magnetic-gating region.
All figures in the paper uses the following parameters:
$\gamma = \SI{2.21e5}{A/m}$, $K = \SI{3.88e4}{A/m}$, $A = \SI{0.328e-11}{\ampere\metre}$, $J = \SI{e6}{A/m}$, and $J' = \SI{0.675e4}{A/m}$, .
For (b, c), the length of the gating region is $l=240~\mathrm{nm}$.
}
\label{fig:capping}
\end{figure*}

{\it Model.}
Instead of real antiferromagnet, for simplicity, we consider a synthetic antiferromagnet (SyAF) consisting of two spatially separated magnetic sub-layers extending in $x$-direction \cite{yang_domain-wall_2015,duine_synthetic_2018}, as depicted in insets of \Figure{fig:capping}(a).
The two sub-layers are coupled antiferromagnetically via the Rudderman-Kittel-Kasuya-Yosida (RKKY) interaction across a metallic spacer layer \cite{parkin_oscillations_1990}. The magnetization dynamics of the SyAF is described by the coupled Landau-Lifshitz-Gilbert (LLG) equations \cite{lan_antiferromagnetic_2017,cheng_antiferromagnetic_2016}
\begin{equation}
\textbf{\.{m}}_j(x,t)
=-\gamma~\textbf{m}_j(x,t)\times\mathbf{H}_j^{\text{eff}}
+\alpha~\textbf{m}_j(x,t)\times\textbf{\.{m}}_j(x,t)
\label{LLGequation}
\end{equation}
where $j=1,2$ denotes the lower and upper sub-layer respectively, $\gamma$ is the gyromagnetic ratio, and $\alpha$ is the Gilbert damping constant. Here $\mathbf{H}_j^{\text{eff}}=Km_j^z\hat{\textbf{z}}+A\nabla^2\textbf{m}_j-J\textbf{m}_{\bar{j}}+\bH_j$ (with $\bar{1}=2, \bar{2}=1$) is the effective magnetic field acting on sub-layer $\textbf{m}_j$, where $K$ is the easy-axis anisotropy along $\hat{\textbf{z}}$, $A$ and $J$ are the intra- and inter-layer exchange coupling coefficients, and $\bH_j$ is the external magnetic field.

The equilibrium magnetization of the two sub-layers $\textbf{m}_{1,2}^0$ points in $\pm\hat{\textbf{z}}$ direction respectively. Upon this collinear magnetic configuration, we separate the static and dynamical components of the sub-layer magnetization as $\textbf{m}_j(x,t) = \textbf{m}_j^0 + \delta\textbf{m}_j(x,t)$, where $\delta\textbf{m}_j = m_j^x\hat{\textbf{x}} + m_j^y\hat{\textbf{y}}$ is the transverse dynamical component of spin wave. By linearizing the LLG equation \Eq{LLGequation} to the leading order of $\tilde{m}_j\equiv m_j^x-im_j^y$, and ignoring the damping term, the spin wave dynamics reduces to:
\begin{equation}
(-1)^j i \pdv{t} \tilde{m}_j
= \gamma \qty(-A\nabla^2+K+J+H_j)\tilde{m}_j+\gamma J\tilde{m}_{\bar{j}}.
\label{LinearizedLLG}
\end{equation}
When the external field vanishes ($H_j=0$), \Eq{LinearizedLLG} hosts two degenerated circularly polarized spin wave modes with dispersions as $\omega_\ssf{L/R}^k =\gamma\sqrt{(Ak^2+K+J)^2-J^2}$ for the left/right circular mode, where $k$ is the wavevector along $\hbx$.

{\it Spin wave polarization manipulation via magnetic gating}. To lift the degeneracy between these two  circular spin wave modes,  we introduce a modified SyAF structure by capping a magnetic gating layer ($\mb_\ssf{G}$) upon the SyAF (see \Figure{fig:capping} (a)). The gating layer is antiferromagnetically coupled to the upper layer of SyAF via RKKY with strength $2J'$. For simplicity, the gating layer magnetization is pinned along $\hbz$ axis with $\mb_\ssf{G} = \pm \hbz$.
The magnetic gating effect of $\mb_\ssf{G}$ on SyAF is introduced via the exchange field $\bH_2=2J'\mb_\ssf{G}$ from the gating layer on the upper sub-layer, while the lower sub-layer is not affected (thus $\bH_1 = 0$). This magnetic gating field modifies the spin wave dispersions in the SyAF to
\begin{equation}
\label{eqn:omega_shift}
\omega_\ssf{L/R}^k = \mp \gamma J' m_\ssf{G}^z + \gamma\sqrt{(Ak^2+K+J- J'm_\ssf{G}^z)^2-J^2} ,
\end{equation}
where $m_\ssf{G}^z=\pm 1$ denotes the magnetization direction of the gating layer.
The modified dispersions in \Eq{eqn:omega_shift} are plotted in \Figure{fig:capping}(a), showing that the magnetic gating either increases or decreases the spin wave gap depending on the gating magnetization direction ($\mb_{\ssf{G}} = \pm\hbz$). In the mean time, the magnetic gating lifts the degeneracy between the left- and right-circular modes due to the preferential coupling between the capping layer and the upper sub-layer of SyAF.
Consequently, the left- and right-circular modes with the same frequency $\omega$ would propagate at different wavevectors $k_\ssf{L,R}$ with $\omega = \omega_\ssf{L}(k_\ssf{L}) = \omega_\ssf{R}(k_\ssf{R})$, which results in a relative phase delay between the right- and left-circular spin wave components.
The magnetic gating effect here is realized via RKKY interaction between the capping ferromagnet and SyAF. Similar gating effect should be possible via the exchange bias effect between ferromagnet and real antiferromagnet \cite{nogues_exchange_1999}, for controlling the spin wave polarizations in the latter.

When a linearly polarized spin wave of frequency $\omega$ is passing through a gating segment of length $l$, this relative phase delay induces a rotation of the linear polarization by angle $\Delta \theta=(k_\ssf{L}-k_\ssf{R})l/2$. In WKB approximation,
\begin{equation}
\label{eqn:delta_theta}
 \Delta\theta\approx  - \frac{\omega  }{   \sqrt{\omega^2 +\gamma^2 J^2} }  \eta(1 - \eta) k_0 l,
\end{equation}
where $\eta= (J'/2Ak_0^2)m_\ssf{G}^z $ denotes the gating efficiency, and $k_0$
is the wavevector without gating. As seen, the rotation direction is explicitly controlled by the gating magnetization direction $m_\ssf{G}^z$ via $\eta$, and the rotation is more efficient for the down-gated case ($m_\ssf{G}^z=-1$).

This magnetic gating induced polarization rotation is a spin wave analog of the Datta-Das spin transistor \cite{datta_electronic_1990}: the spin wave polarization rotates in opposite directions when the magnetic gating reverses or $m_\ssf{G}^z$ changes sign.
This spin wave Datta-Das field transistor is of purely magnetic nature, thus is different from a spin wave field transistor proposed by Cheng {\it et. al.} using an electrical gating of the Dzyaloshinnski-Moriya interaction (DMI) \cite{cheng_antiferromagnetic_2016}.
\Figure{fig:capping}(b) shows a micromagnetic simulation of the magnetic gating effect on a linear-$y$ (\ang{90}) polarized spin wave of frequency $f = \omega/2\pi = \SI{11.5}{GHz}$.
It is seen that, as the spin wave passing across the magnetic gating region, its linear polarization steadily rotates (counter-)clockwise, and finally acquires a rotation angle of $\Delta\theta = $ +\ang{26}/\ang{-19} for down-/up-gating.
The frequency dependence of the rotation angle $\Delta \theta$ is shown in \Figure{fig:capping}(c), agreeing with the scattering calculations and the WKB expression in \Eq{eqn:delta_theta}.

The magnetic gating provides the capability of converting the information from static magnetic domains to polarized spin waves.
To form a complete interchanging scheme between static magnetic domains and dynamical spin waves, one needs an extra ingredient of manipulating magnetic domains via polarized spin wave.
It is known that, in the presence of DMI (which usually naturally exists in SyAF), the antiferromagnetic domain wall reflects spin waves differently according to its linear polarization \cite{lan_antiferromagnetic_2017}.
The associated back reaction of the spin wave reflection is that the domain wall is pushed forward by the reflected polarization ($\hby$ polarization here) but stays still for the transmitting polarization ($\hbx$ polarization) \cite{yu_polarization-selective_2018}.
With these two complementary ingredients, we are ready to construct logic gate of purely magnetic nature.

\begin{figure}[t]
\centering
\includegraphics[width=0.5\textwidth]{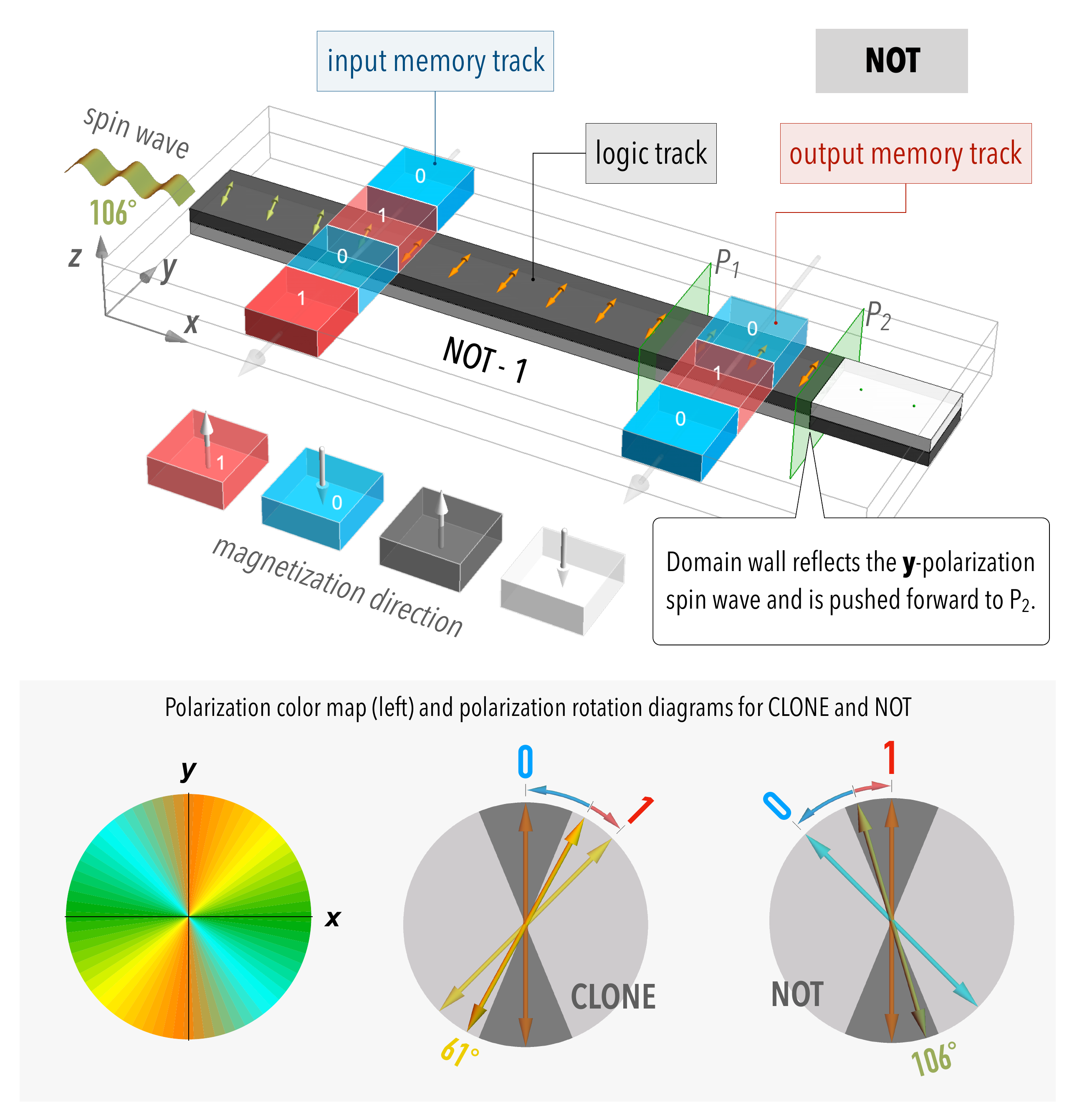}
\caption{A magnetic unary gate consists of an input and an output memory track (magnetic racetrack memory) bridged by a logic track (a SyAF wire with a domain wall). The racetrack memories can be driven by electric current to shift desired bit into (out of) the logic track. This unary gate realizes the NOT function for $\theta_i = \ang{106}$ (as shown), but the CLONE function for $\theta_i =  \ang{61}$ (not shown). Bottom: The color map of the linear spin wave polarization (left), the polarization rotation diagram for the CLONE gate (middle) and the NOT gate (right). The shaded area indicates the polarization angles which can push the domain wall to P$_2$. }
\label{fig:unary_gate}
\end{figure}

{\it Unary gates.}
The simplest logic gate of all is the NOT gate. A magnetic NOT gate, shown in \Figure{fig:unary_gate}(a), consists of two magnetic racetrack memories bridged by a logic track made of SyAF. All adjacent layers are coupled antiferromagnetically via RKKY interaction.
The input and output bits are stored as magnetic domains in the racetrack memory with spin up/down for bit-1/0.
In the logic track, an antiferromagnetic domain wall is located close to the output track. The linearly polarized spin wave is injected from the left and propagates along the logic track toward the domain wall.
Without driving spin waves, the domain wall stays at or relaxes back to position P$_1$, by anisotropy gradient for example \cite{franken_shift_2012}. Under influence of driving spin waves, the domain wall may or may not be pushed up to position P$_2$, depending on the polarization of the incoming spin wave.

The magnetization in the input memory track acts as a magnetic gate, modulating the spin wave polarization running in the logic track according to the input bit: the bit-0 (1) corresponds to up (down)-gating.
For a NOT gate, the polarization of injected spin waves (of frequency $f=\SI{11}{GHz}$) is chosen as $\theta_i=\ang{106}$, and the length of the gating region (equals to the width of the input track) is $l = 240~\si{nm}$ such that an overall rotation angle of $\Delta\theta_0=+\ang{29}$ for bit-0 or $\Delta\theta_1=\ang{-16}$ for bit-1, and $\Delta\theta_0 - \Delta\theta_1=\ang{45}$.
\footnote{The rotation angle is reduced because of partial reflection of $\hbx$ polarization at $f=\SI{11}{GHz}$.}
Via the magnetic gating, the bit information stored in the input track is read out and encoded into the rotated spin wave polarization in the logic track: $\theta_{0/1} = \theta_i+\Delta\theta_{0/1} = \ang{135}/\ang{90}$ for bit-0/1, as shown in  the polarization rotation diagram for NOT in \Figure{fig:unary_gate}.

The following information processing is realized using the polarization-selective spin wave driven domain wall motion mentioned above and detailed in Ref. [\onlinecite{yu_polarization-selective_2018}]. More specifically for the NOT gate, the modulated spin wave, after being gated by the input bit-1, is polarized along $\hby$ (\ang{90}), therefore the spin wave is completely reflected and this transfers enough momentum to push the domain wall position P$_1$ across the output track to position P$_2$. On the other hand, the modulated spin wave gated by input bit-0 has \ang{135}-polarization, far away from $\hby$-polarization, the reflection is small, and the domain wall stays at or returns to P$_1$.

The output is written to the output memory via magnetic imprinting: because the magnetizations in the upper sub-layer of SyAF and the output track are always opposite, when the domain wall is at P$_1$, the upper sub-layer magnetization of SyAF is up, thus the magnetization in the output track is down, corresponding to bit-0. The overall results are: for the input bit-1 (0), the modulated spin wave pushes the domain wall to P$_1$ (P$_2$), thus bit-0 (1) is imprinted to the output track, realizing a NOT gate with $1\to 0$ and $0\to 1$. Multiple cycles of NOT operations are  simulated using micromagnetic simulations (see Supplementary Movies).

A unique advantage of this magnetic logic is its extreme configurability. As shown above, the gate structure in \Figure{fig:unary_gate} functions as a NOT gate when injecting spin wave with polarization $\theta_i = \ang{106}$. Merely by changing the injection polarization to $\theta_i = \ang{61}$, this unary gate manifests itself as a CLONE gate, \textit{i.e.} cloning the input bit to the output memory.

{\it Binary gates.} For the binary gates, there are three possible input combinations: $00$, $01/10$, $11$ and two possible outputs: $0$ or $1$, therefore there are in total $2^3=8$ distinct gates. These include six non-trivial gates: OR, AND, NOR, NAND, XOR, XNOR, and two trivial gates that map all input combinations unanimously to either $0$ or $1$ (called ZERO and UNITY here).

\begin{figure}[t]
\centering
\includegraphics[width=0.5\textwidth]{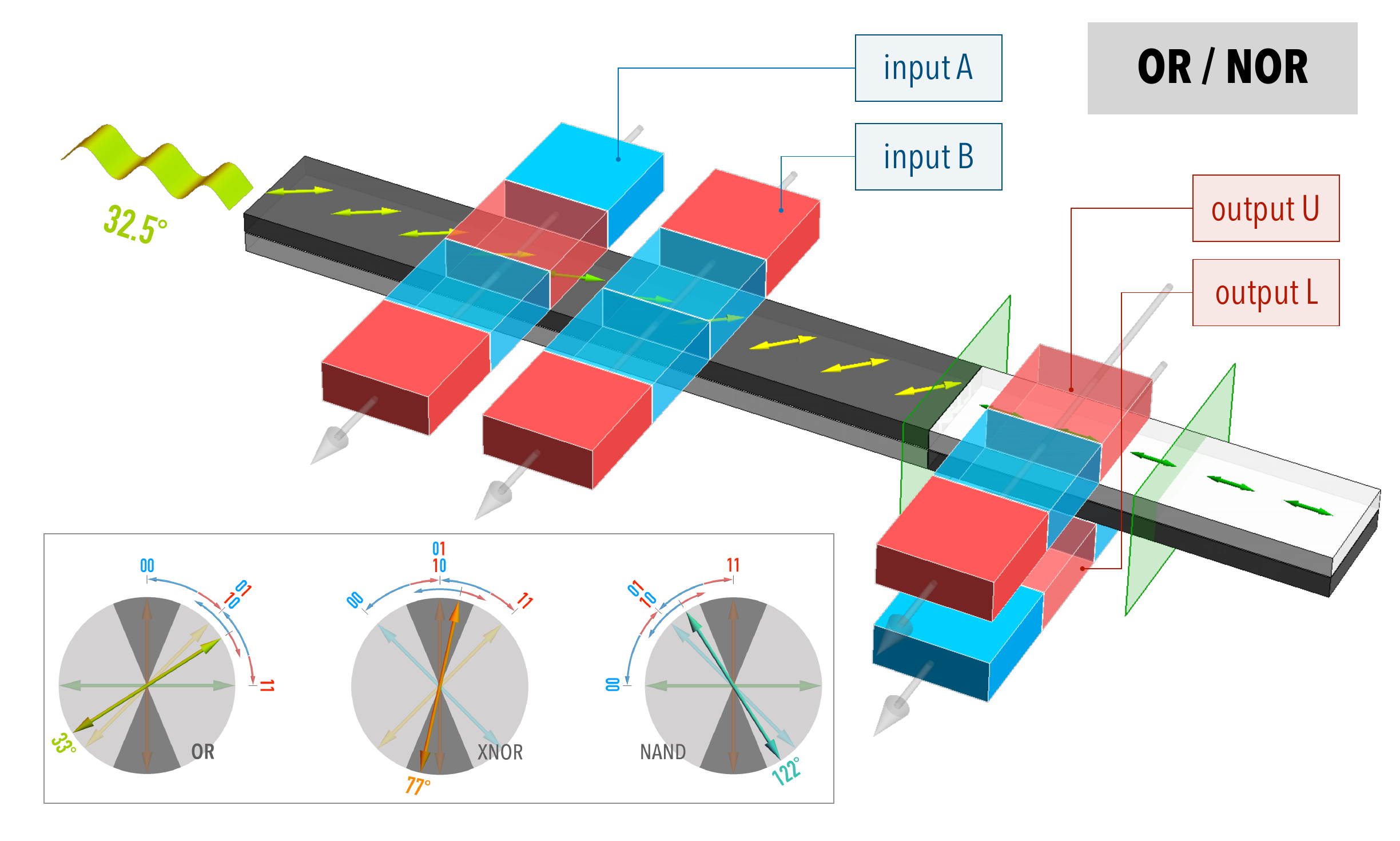}
\caption{A magnetic binary gate consists of two input racetrack memories and (at least) one output racetrack, bridged by a logic track. Inset: the polarization rotation diagram for the OR/XNOR/NAND gate with the initial polarization angle \ang{32}/\ang{77}/\ang{122} . }
\label{fig:binary_gate}
\end{figure}

The binary gates can be extended from the unary gate by adding one more input track as depicted in \Figure{fig:binary_gate}, which has now two input memory tracks (A and B) and two output tracks (U and L), bridged by an SyAF logic track. Since all adjacent layers are coupled antiferromagnetically, the upper/lower output (U/L) always yield opposite results, {\it e.g.} if output U is logic OR, then output L gives logic NOR.

With two input tracks, the injected spin wave is gated by two consecutive magnetic gating segments.
Because the polarization rotation caused by bit-0 and bit-1 are the opposite, the overall rotation can be either enhanced when the inputs are the same (00 or 11), or reduced when the inputs are different (01 or 10).
For three different input combinations, the spin wave polarization would rotate by $\Delta\theta_{00}=2\Delta\theta_0=+ \ang{58}$ for input 00, $\Delta\theta_{11}=2\Delta\theta_1=\ang{-32}$ for input 11, and $\Delta\theta_{01}=\Delta\theta_{10}=\Delta\theta_0+\Delta\theta_1=+\ang{13}$ for input 01 or 10, each separated by exactly \ang{45}.

When the injected spin wave has polarization angle $\theta_i=+\ang{32}$ (as shown in \Figure{fig:binary_gate}), the modulated polarization after gating becomes $\theta_{00} = \ang{90}$ for input 00, $\theta_{11} = \ang{0}$ for input 11, and $\theta_{01}=\theta_{10} = \ang{45}$ for input 01 or 10, respectively. Since only $\theta_{00}$ has large enough $\hby$-component (within the shaded area in the rotation diagrams in \Figure{fig:binary_gate}), the domain wall will move across the output tracks to P$_2$ only for input 00, and the moved domain wall imprints bit-0 to the output U track: $00\to 0$. For all other inputs, the domain wall stays at P$_1$, which writes bit-1 to the output U track: $01/10/11\to 1$.
Therefore, the binary gate shown in \Figure{fig:binary_gate}, with injecting polarization $\theta_i=+\ang{32}$, realizes an OR gate in the output U track, and in the mean time a NOR gate in the output L.

Similar to the unary gate, the binary gate structure in \Figure{fig:binary_gate} can also serve as multiple different gates by simply altering the polarization of the injected spin waves. Based on the polarization rotation diagrams in \Figure{fig:binary_gate}, the gates OR/XNOR/NAND (NOR/XOR/AND) are realized in the output U (L) track when the injected spin wave are polarized along $\ang{32}/\ang{77}/\ang{122}$ (see Supplementary Movies).

\begin{figure}[t]
\centering
\includegraphics[width=0.488\textwidth]{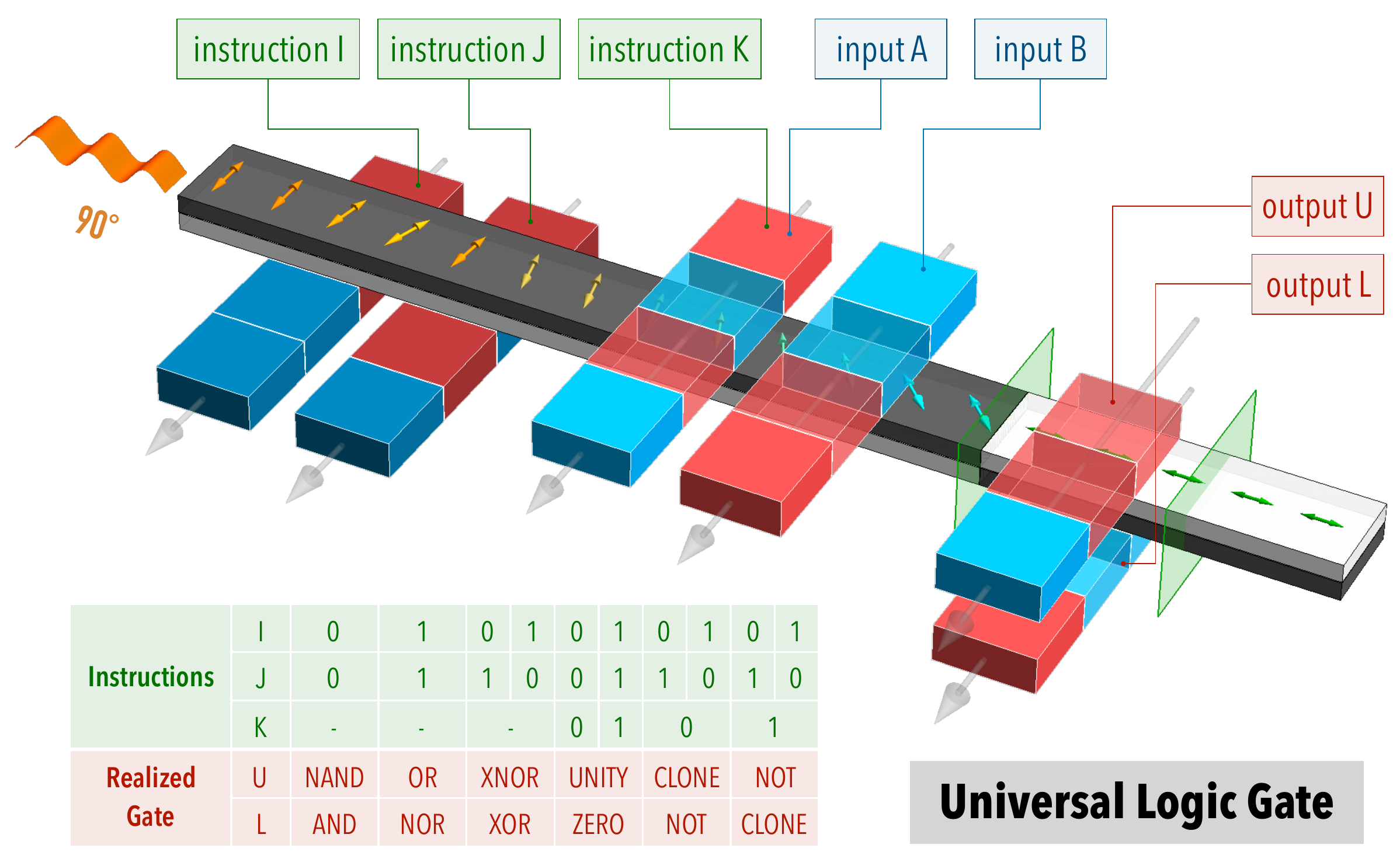}
\caption{The universal magnetic logic gate with two instruction tracks (I, J) can realize all unary and binary logic gates with the fixed \ang{90} spin wave polarization injection. Inset: The instruction table for realizing different gates. }
\label{fig:universal_gate}
\end{figure}

{\it Universal gate.} Although the unary and binary gates above can realize different gate functions by varying injected spin wave polarizations, it is inconvenient to do so. To avoid this complication, we propose a universal logic gate as shown in \Figure{fig:universal_gate}, which uses two more instruction tracks (I, J). According to the bits in the instruction tracks, the \ang{90} polarized spin wave can be pre-processed into the desired $\ang{32}/\ang{77}/\ang{122}$ polarization. For instance, when the instruction bits are IJ $= 10$, the polarization after the instruction tracks is $\ang{77}$, realizing XNOR/XOR for the output U/L.
This structure can also function as unary gate by using input the A as the third instruction K.
For instance, when  IJK $=010$, it is a CLONE/NOT gate from input B to output U/L.

According to the instruction table in \Figure{fig:universal_gate}, this structure realizes all ten unary and binary logic gates, therefore we call it a universal logic gate. In other words, a magnetic logic gate using only one logic track can realize multiple functionalities which usually require a dozen of conventional electronic gates. Micromagnetic simulations confirm the functioning of the universal logic gate (see Supplementary Movies).

{\it Spin wave arithmetic logic unit (ALU).}

\begin{figure*}[t]
\centering
\includegraphics[width=0.95\textwidth]{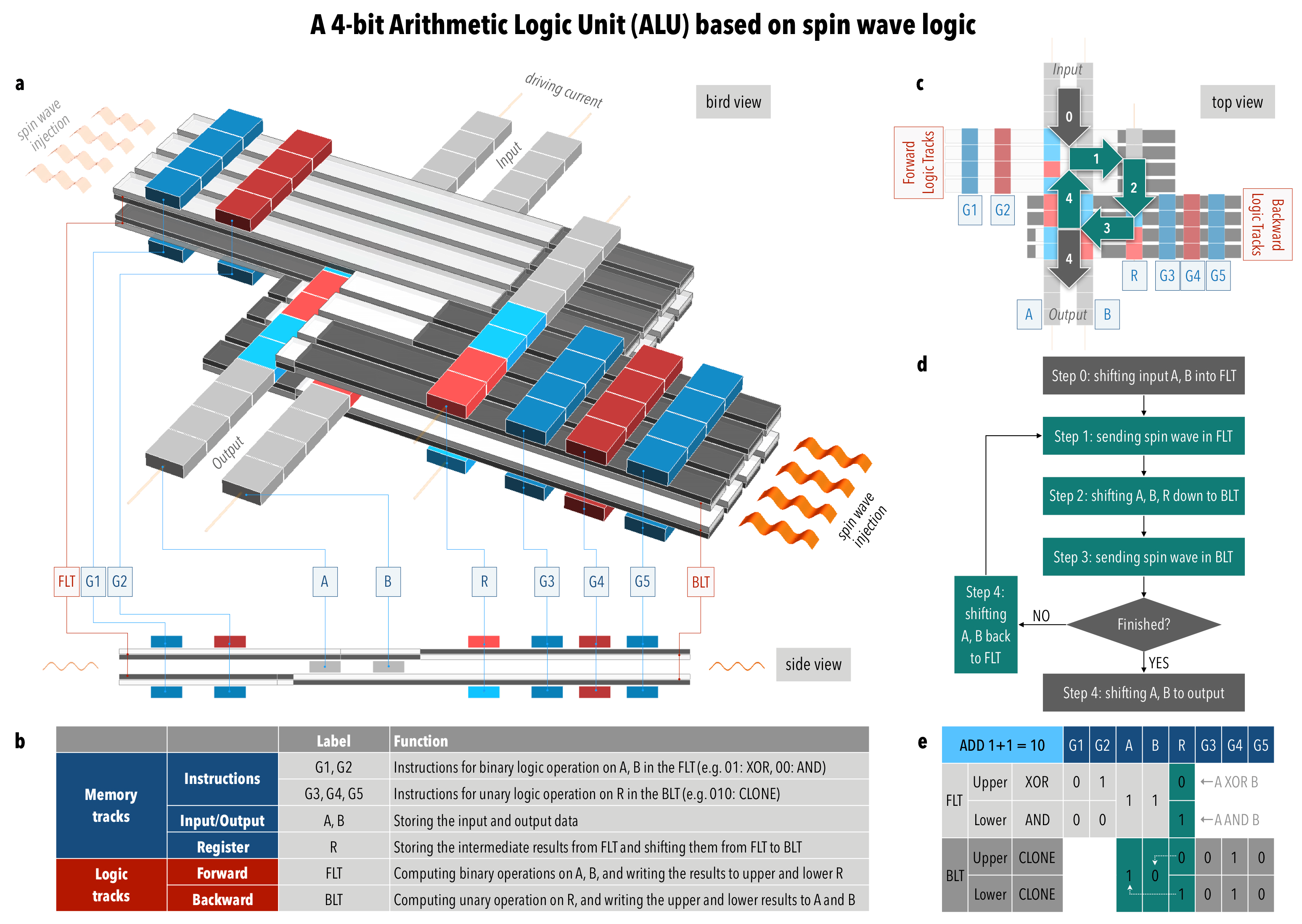}
\caption{The 4-bit Arithmetic Logic Unit based on the universal spin wave logic.  a. The ALU consists of 10 racetrack memory (double-)tracks interweaving with 8 pairs of logic tracks. b. The function of each memory and logic track. c. The top view of the ALU and its operating order. d. The flow chart for the working process of the spin wave-based ALU. e. The simplified version of 1-bit ALU working out the 1-bit sum of 1+1 = 10.}
\end{figure*}

The Arithmetic Logic Unit (ALU) is the core component of a Central Processing Unit (CPU) in modern computers. The tasks of an ALU is to carry out some of the most basic logic and arithmetic operations, such as AND, OR, XOR, SHIFT LEFT/RIGHT, ADD, etc. In conventional computers, a partial-functional 4-bit ALU typically needs over one hundred NAND gates \cite{scott_but_2009}. In contrast, constructing a full-functional 4-bit ALU based on the spin wave logic, as presented below, only needs 16 universal gates. The spin wave ALU can not only carry out all unary and binary logic operations, but also the advanced arithmetic operations such as SHIFT LEFT/RIGHT and ADD/SUBTRACT with carry-in and -out.
The construction of the spin wave-based ALU is depicted in Figure 4a, which interweaves 8 memory (double-)tracks with 16 SyAF logic tracks in an orthogonal manner. The logic tracks are partitioned into Forward Logic Tracks (FLT) and Backward Logic Tracks (BLT), each consists of 4 pairs of upper and lower SyAF logic tracks. Each pair of upper and lower logic tracks act as two distinct universal logic gates, whose instructions are stored in track G1-G5. The upper and lower FLT share the input tracks A and B placing between the upper and lower logic tracks, and writes the outputs to the upper/lower Register track R. The BLT are unary gates, which CLONE or NOT the bits in the upper/lower R track and write results to B/A. The functions of all tracks are listed in Figure 4b.
In the ALU in Figure 4a, the upper/lower instructions on G1, G2 are 01/00 (XOR/AND), therefore the upper/lower FLT tracks perform XOR/AND operations on input A, B. The instructions in G3-G5 are 010 (CLONE) for both upper and lower BLT, therefore the BLT simply CLONE upper/lower R to track B/A. The full operating cycle of the ALU is decomposed into five steps as shown in Figure 4c and 4d: 0). The 4-bit data to be processed is shifted from the input port into the FLT region by current-driven racetrack memory in track A and B; 1). Injecting $90^\circ$-polarization spin wave into the FLT from the left, the FLT carries out logic operation on input A and B according to the instructions in G1-G2 and writes the results into Register R; 2). Shifting A, B, R by four positions from FLT to the BLT; 3). By injecting spin wave into the BLT from the right, the BLT CLONEs the results stored in R back into track A, B, thus overwriting original inputs stored in A, B; 4). If the output is final (as for logic operations), the track A and B are shifted to the output port. Otherwise, the track A and B are shifted back to the FLT for the next cycle (as for addition or subtraction). A simplified 1-bit version is shown in the table of Figure 4e: In FLT, the upper/lower track perform XOR/AND on input A = 1, B = 1, thus the upper/lower R track is 0/1 after step 1. In step 2, the 0/1 in R is shifted into BLT. In step 3, the 0/1 in R is CLONED to B/A, thus A = 1, B = 0. After this full cycle, the ALU carried out the addition of A and B: 1 + 1 = 10. An example for adding two 4-bit numbers with carry-in and carry-out can be found in the Supplementary Materials.

{\it Discussion \& Conclusions.}
The magnetic logic introduced in this paper has several unique features.
First, the architecture is of purely magnetic nature, where both the data storage and processing are achieved using magnetic elements.
Second, since the all inputs and outputs are stored in the non-volatile racetrack memories, the magnetic logic naturally realizes non-volatile memory-to-memory (or in-memory) computing. Because of this nature, the relatively short lifetime of spin wave is not a serious issue as long as it can sustain to accomplish one single operation, after which the result is stored.
Third, the extreme reconfigurability or the universality comes from the double-threshold nonlinearity of polarized spin waves (See Supplementary Materials). As a result, the hardware itself is not only programable and can even evolve on real time, much more flexible than the field programable gate arrays (FPGA). This capability makes it possible to realize evolvable hardware based on magnetic logic, while fulfilling essential characteristics for scalable computing \cite{behin-aein_proposal_2010} (See Supplementary Materials).

In conclusion, we introduce the concept of magnetic gating on spin wave polarizaiton, which gives rise to a purely magnetic analog of the Datta-Das spin transistor. Based on this magnetic gating effect, we proposed the a universal logic gate of purely magnetic nature, with data stored in the the static magnetic textures (domains) and processed by its dynamical excitations (polarized spin wave).
Because of its non-volatility and universality, this magnetic logic concept provides new designing principles for in-memory processing.

{\it Acknowledgements.} This work was supported by the National Natural Science Foundation of China (No. 11722430, No. 11847202).
W.Y. is also supported by the China Postdoctoral Science Foundation (No. 2018M641906).

\bibliography{refs}

\end{document}